\documentclass[sigconf]{acmart}

\AtBeginDocument{%
  }

\usepackage{amsfonts}  
\usepackage{multirow}
\usepackage{array}
\usepackage{CJKutf8}   

\begin{document}

\title[Character-Level Prosody Control for Mandarin TTS]{%
Character-Level Prosody Control for Mandarin Speech Synthesis:\\
An Architecture-Preserving Approach}


\author{Diankun An}
\affiliation{%
 \institution{JD.com, Inc.}
 \city{Beijing}
 \country{China}}
\email{andiankun.1@jd.com}

\author{Lu Ma}
\authornote{Corresponding author. Lu Ma conceived the idea and demonstrated the feasibility of the proposed method through small-scale experiments. Diankun An conducted large-scale experiments that further validated the feasibility and effectiveness of the method and provided detailed experimental results. All authors contributed to the discussion and writing of the manuscript.}
\correspondingauthor
\authornotemark[1]
\affiliation{%
 \institution{JD.com, Inc.}
 \city{Beijing}
 \country{China}}
\email{iamroad@163.com}

\author{Yongliang Wang}
\affiliation{%
 \institution{JD.com, Inc.}
 \city{Beijing}
 \country{China}}
\email{wangyongliang1@jd.com}

\author{Zhewen Su}
\affiliation{%
 \institution{JD.com, Inc.}
 \city{Beijing}
 \country{China}}
\email{suzhewen.1@jd.com}

\author{Xin Tian}
\affiliation{%
 \institution{JD.com, Inc.}
 \city{Beijing}
 \country{China}}
\email{tianxin84@jd.com}

\renewcommand{\shortauthors}{Lu Ma et al.}

\begin{abstract}
This paper addresses fine-grained prosody control for large-language-model
(LLM) based text-to-speech (TTS) synthesis of \textbf{Mandarin Chinese},
motivated by multimedia content-creation scenarios---film dubbing, audiobook
production, and interactive voice interfaces---in which prosodic adjustments
must be placed on specific characters rather than applied uniformly to whole
utterances. We present a data-driven framework in which fine-grained prosodic
markers (pitch, duration, energy, and pause) are automatically extracted from
raw audio at scale and embedded into the text stream as out-of-vocabulary
Unicode symbols. The LLM of an existing TTS system is then fine-tuned to
interpret these markers while its architecture remains unchanged, yielding
character-level temporal control that complements existing utterance-level
style control. An optional text-LLM front-end can translate natural-language
instructions into marker sequences. Experiments on 2.15M Mandarin utterances (5,858
hours) with CosyVoice2 as the backbone demonstrate reliable control of
character duration, pitch, pause duration, and global speaking rate, with
perceptual quality essentially preserved. A cross-tokenizer analysis further
indicates that the ASR-optimized tokenizer of the backbone attenuates
prosodic variation relative to reconstruction-oriented alternatives---an
observation we report as hypothesis-generating evidence for prosody-sensitive
system design.
\end{abstract}





\settopmatter{printacmref=false}
\setcopyright{none}
\renewcommand\footnotetextcopyrightpermission[1]{}

\maketitle

\pagestyle{plain}

\section{Introduction}
\label{sec:intro}

The proliferation of AI-generated content (AIGC) has created strong demand
for controllable speech synthesis in multimedia production. Film dubbing
requires precise emphasis on specific words, audiobook narration requires
rhythmic variation to sustain listener engagement, and interactive gaming and
virtual-reality applications require prosodic adjustments in response to user
actions. While recent LLM-based TTS systems (e.g.,
CosyVoice~\cite{du24cosyvoice}, Spark-TTS~\cite{wang25sparktts},
Seed-TTS~\cite{zeng2024seedtts}) produce highly natural speech with
utterance-level style control, they lack \emph{surgical} prosodic
manipulation at the character level. \textbf{Scope.} This work focuses on
Mandarin Chinese: our marker inventory, alignment pipeline, and all
experiments are designed for and validated on Mandarin speech, in which
characters constitute a natural prosodic control unit. Generalization to
other languages is discussed as future work (Section~\ref{sec:discussion}).

Existing control mechanisms fall into two categories, each with distinct
limitations~\cite{xie2025towards}. \textbf{Utterance-level control} via
natural-language descriptions (e.g., ``speak sadly'') or reference speech
provides global style transfer but cannot specify \emph{which} syllable
should rise in pitch or \emph{where} exactly to
pause~\cite{yang2024instructtts,leng24prompttts}. \textbf{Fine-grained
symbolic control} via SSML, ToBI~\cite{silverman1992tobi}, or learned
word/phoneme-level tags~\cite{korotkova2024wordlevel,guo2022unsupervised,mohan2021ctrlp,vioni2021prosodic}
offers local precision but typically requires expert annotation,
language-specific rules, or modification of the acoustic model, which limits
scalability and prevents direct application to pre-trained LLM-based TTS
systems.

We bridge this gap with a data-driven framework for character-level prosody
control of Mandarin TTS. Prosodic patterns are automatically extracted from
raw audio via forced alignment and statistical analysis and embedded into
text as out-of-vocabulary Unicode markers; the LLM of an existing TTS system
then learns to interpret these markers through fine-tuning alone. We note
that automatic prosody annotation from audio--text pairs has a substantial
history, from HMM-era simultaneous modeling of spectrum, F0, and
duration~\cite{yoshimura1999simultaneous} to recent pre-trained-model-based
annotation~\cite{dai2022automatic,tilson2024toward}; our contribution is
therefore \emph{not} automatic extraction per se, but (i) a character-level
Unicode marker design tailored to Mandarin and (ii) the finding that a
frozen-architecture TTS-LLM can be taught to follow such markers,
establishing an interpretable, text-side control channel. An optional
front-end that translates natural-language instructions into marker sequences
is summarized in Section~\ref{subsec:frontend}.

This work makes three contributions:
\begin{itemize}
    \item \textbf{Automated character-level prosody markup at scale.} A
    pipeline that extracts pitch, duration, energy, and pause markers from
    raw audio without manual annotation, demonstrated on 2.15M Mandarin
    utterances (5,858 hours).
    \item \textbf{Architecture-preserving marker interpretation.} Evidence
    that the LLM of an existing TTS system (CosyVoice2) can be fine-tuned to
    realize discrete character-level markers as acoustic variation without
    any structural change, providing control that is complementary to
    utterance-level style conditioning.
    \item \textbf{Cross-tokenizer sensitivity analysis.} Measurements
    suggesting that ASR-optimized tokenizers attenuate prosodic perturbations
    relative to reconstruction-oriented ones, reported as
    hypothesis-generating evidence for prosody-sensitive tokenizer design.
\end{itemize}

\textbf{Research objective.} The goal of this work is \emph{not} to
outperform existing systems in overall naturalness, but to demonstrate that
character-level prosodic markers can be automatically obtained and reliably
interpreted by an existing TTS-LLM. We treat \emph{naturalness} and
\emph{realism} as distinct perceptual dimensions: human speech frequently
exhibits intentional prosodic deviations (e.g., irregular lengthening in
nervous speech, deliberate pitch exaggeration in dramatic performance) that
depart from neutral naturalness yet are realistic and desirable for
expressive content. Our evaluation therefore focuses on feasibility,
controllability, and multi-dimensional perceptual quality, and the framework
should be viewed as a complement to utterance-level control methods.

\section{Related Work}
\label{sec:related}

\subsection{Utterance-Level Prosody Control}

Modern LLM-based TTS systems primarily offer utterance-level control.
PromptTTS2~\cite{leng24prompttts,shimizu24prompttts},
PromptStyle~\cite{liu23promptstyle}, and
InstructTTS~\cite{yang2024instructtts} use natural-language descriptions for
global style, which inherently lacks temporal precision for intra-sentence
variation. Reference-speech methods (Seed-TTS~\cite{zeng2024seedtts},
Mega-TTS~2~\cite{jiang24megatts2}, NaturalSpeech~2/3~\cite{shen24naturalspeech2,ju24naturalspeech3},
Style Tokens~\cite{wang2018style}, EmoCtrl-TTS~\cite{wu2024laugh}) transfer
acoustic style holistically. These approaches are complementary to ours: they
set global style, while our markers adjust local prosody.

\subsection{Fine-Grained Symbolic and Markup Control}
\label{subsec:symbolic}

SSML and ToBI~\cite{silverman1992tobi} provide character- or phoneme-level
control through explicit markup, but require expert annotation and
language-specific rules. Several neural approaches reduce this cost:
Ctrl-P~\cite{mohan2021ctrlp} enables temporal control of prosodic variation,
Vioni et al.~\cite{vioni2021prosodic} derive phoneme-level control from
prosodic clustering, Guo et al.~\cite{guo2022unsupervised} perform
unsupervised word-level prosody tagging, and Korotkova et
al.~\cite{korotkova2024wordlevel} introduce word-level text markup trained
with automatic annotation. FastSpeech~2~\cite{ren2020fastspeech} predicts
phoneme-level duration and pitch directly, but requires modifying the
acoustic model. Our work differs in scope and mechanism: markers operate on
Mandarin \emph{characters}, are inserted as out-of-vocabulary Unicode symbols
into the text stream, and are realized by an off-the-shelf LLM-based TTS
whose architecture is left intact.

\subsection{Automatic Prosody Annotation}

Extracting prosodic structure from speech has been studied since HMM-based
synthesis jointly modeled spectrum, F0, and
duration~\cite{yoshimura1999simultaneous}. More recently, Dai et
al.~\cite{dai2022automatic} annotated prosody automatically with a
pre-trained text--speech model, and Tilson et
al.~\cite{tilson2024toward} moved toward real-time word-based prosody
recognition. We build on this line of work: our extraction pipeline is an
engineering contribution aimed at scale and Mandarin character granularity,
rather than a claim that automatic prosody extraction is itself new.

\subsection{LLM-Based Synthesis and Inline Control Tokens}

VALL-E~\cite{wang2023neural} and subsequent systems
(CosyVoice~\cite{du24cosyvoice}, Spark-TTS~\cite{wang25sparktts},
IndexTTS~\cite{deng25indextts}, MsEmoTTS~\cite{lei22msemotts}) treat TTS as
conditional language modeling but lack fine-grained temporal control
mechanisms. BatonVoice~\cite{wang25batonvoice} uses an external LLM as a
``conductor'' producing phrase-level vocal plans, requiring a specialized TTS
trained from scratch; NVSpeech~\cite{liao2025nvspeech} inserts discrete
inline tokens for paralinguistic events (e.g., \texttt{[Laughter]}), a form
of content augmentation that expands the vocabulary and sequence. Our markers
instead modulate \emph{existing} lexical content at the acoustic-attribute
level, without introducing new lexical items or task-specific vocabulary
entries; the approaches are complementary.

\section{Methodology}
\label{sec:method}

\begin{figure}[t]
\centering
\includegraphics[width=\linewidth]{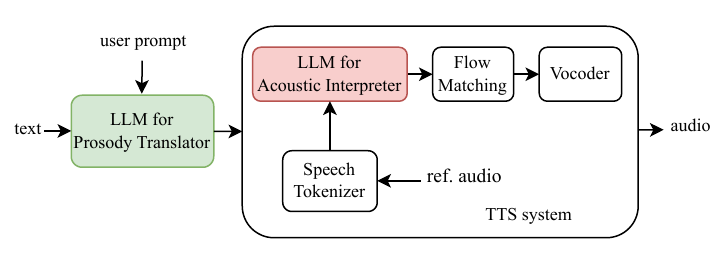}
\caption{Inference-time view of the framework. At inference, marked text is
authored directly or generated by the optional instruction front-end; the
offline extraction pipeline (Figure~\ref{fig:modelling}) supplies the marked
training data. The base model architecture is unchanged.}
\Description{Block diagram of the inference path: marked text enters the
fine-tuned TTS language model, which outputs acoustic tokens decoded to
speech.}
\label{fig:system}
\end{figure}

The framework comprises (1) an \emph{automated extraction pipeline} that
converts raw audio--text pairs into prosodically marked text
(Figure~\ref{fig:modelling}), and (2) an \emph{acoustic interpreter}: the
LLM of an existing TTS system, fine-tuned to render the markers into acoustic
tokens (Figure~\ref{fig:system}). An optional natural-language front-end is
described in Section~\ref{subsec:frontend}.

\subsection{Automated Character-Level Prosody Extraction}
\label{subsec:modelling}

\begin{figure*}[t]
\centering
\includegraphics[width=0.85\textwidth]{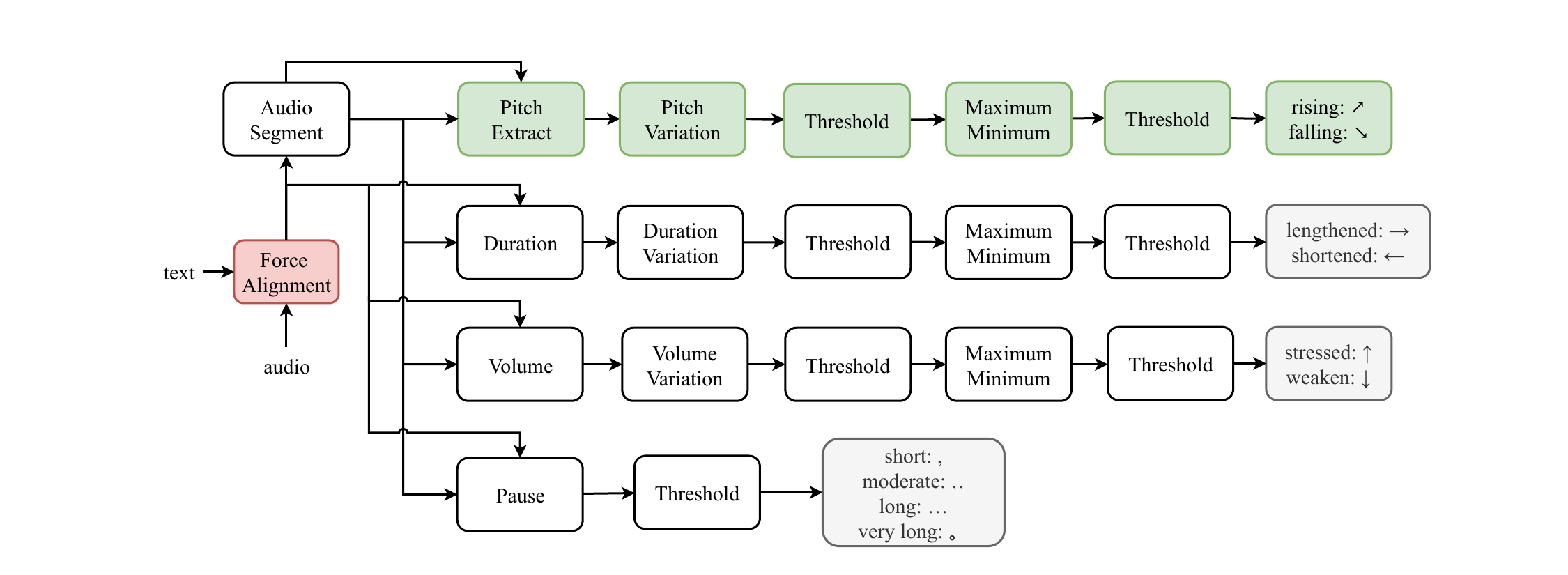}
\caption{Character-level prosody extraction. Forced alignment provides
temporal boundaries for feature extraction and marker insertion.}
\Description{Flowchart of the extraction pipeline: audio and text are force
aligned by two systems whose boundary estimates are fused; per-character
pitch, duration, energy, and inter-phrase pause features are then computed
and converted to Unicode markers inserted around characters.}
\label{fig:modelling}
\end{figure*}

The pipeline (Figure~\ref{fig:modelling}) extracts prosodic markers via
statistical analysis of aligned audio--text pairs, replacing manual markup
with automatic, data-driven annotation.

\textbf{Forced alignment and segmentation.} We combine
MFA~\cite{mcauliffe17_interspeech} (acoustic precision) and
WhisperX~\cite{bain23_interspeech} (robustness to spontaneous speech) to
obtain frame-level character timestamps. Samples for which the two systems
disagree by more than 150\,ms are discarded. For retained samples, boundary
estimates are fused by confidence-weighted averaging:
\begin{equation}
\hat{t} = \frac{w_{\text{W}}\, t_{\text{W}} + w_{\text{M}}\, t_{\text{M}}}{w_{\text{W}} + w_{\text{M}}}
\label{eq:fusion}
\end{equation}
where $w_{\text{W}}$ and $w_{\text{M}}$ are normalized confidence scores
derived from WhisperX's CTC posterior probabilities~\cite{graves2006ctc} and
MFA's acoustic log-likelihoods, respectively. Long utterances are then
segmented into phrases at the fused boundaries.

\textbf{Prosodic feature extraction.} For each phrase and character we
extract:
\begin{itemize}
    \item \textbf{Pause duration.} Inter-phrase pauses are quantized into
    four levels (\emph{short}, \emph{moderate}, \emph{long}, \emph{very
    long}); the corresponding duration ranges are given in
    Table~\ref{tab:pause_duration}.
    \item \textbf{Character duration.} Per-character durations are derived
    from alignment timestamps. Thresholds are applied phrase-locally:
    annotation is performed only within phrases whose duration variance
    exceeds a threshold, where characters above the upper boundary
    ($t>430$\,ms) are labeled \emph{lengthened} and those within the lower
    band ($30<t<50$\,ms) \emph{shortened}. The boundaries are corpus-derived
    constants, corresponding to the 95th percentile (430\,ms) and the
    lowest 5\% band (30--50\,ms) of the per-character duration distribution
    of the aligned corpus.
    \item \textbf{Character energy.} Analogously, within high-variance
    phrases, characters with local maximum and minimum energy are labeled
    \emph{stressed} and \emph{weak}, respectively.
    \item \textbf{Pitch (F0).} F0 is estimated with
    RMVPE~\cite{wei23rmvpe}. Within a phrase, a character whose mean
    F0 deviates from the \emph{phrase mean} by more than
    $\pm100$\,Hz is labeled \emph{rising} or \emph{falling}, provided
    that (i) the character's \emph{internal} F0 range stays below
    35\,Hz---so the label reflects a clean level shift relative to
    the phrase rather than intra-character movement---and (ii) its
    duration exceeds 150\,ms. We use absolute Hz thresholds because
    marker granularity is defined over the corpus distribution;
    relative (log-scale/semitone) thresholds are a natural refinement
    left to future work.
\end{itemize}

\begin{table}[t]
\centering
\small
\caption{Prosodic marker definitions.}
\label{tab:markers}
\begin{tabular}{llcc}
\toprule
\textbf{Parameter} & \textbf{Category} & \textbf{Symbol} & \textbf{Unicode} \\
\midrule
\multirow{2}{*}{Pitch} & Rising & $\nearrow$ & U+2197 \\
 & Falling & $\searrow$ & U+2198 \\
\midrule
\multirow{2}{*}{Energy} & Stressed & $\uparrow$ & U+2191 \\
 & Weak & $\downarrow$ & U+2193 \\
\midrule
\multirow{2}{*}{Duration} & Lengthened & $\rightarrow$ & U+2192 \\
 & Shortened & $\leftarrow$ & U+2190 \\
\midrule
\multirow{4}{*}{Pause} & Short & , & U+201A \\
 & Moderate & $\cdot\cdot$ & U+2025 \\
 & Long & … & U+2026 \\
 & Very long & $\circ$ & U+3002 \\
\bottomrule
\end{tabular}
\end{table}

\textbf{Global speaking rate.} The mean character duration of each utterance
is quantized into five levels (\emph{very slow}, \emph{slow}, \emph{normal},
\emph{fast}, \emph{very fast}) using the 5th/20th/80th/95th percentiles of
the corpus distribution; the resulting boundaries are reported in
Section~\ref{subsec:setup}.

\textbf{Marker insertion.} Markers (Table~\ref{tab:markers}) are inserted
immediately adjacent to the character they modify---as prefixes for pitch
and energy, and as suffixes for duration and pause---and the global rate
prompt precedes the text, separated by a special token
\texttt{<|endofprompt|>}. This transformation preserves the original lexical
content while overlaying prosodic instructions (Table~\ref{tab:example}).

\begin{table}[t]
\centering
\renewcommand{\arraystretch}{1.4}
\caption{Example of prosodic annotation (Mandarin).}
\label{tab:example}
\begin{tabular}{lp{0.30\textwidth}}
\toprule
\textbf{Category} & \textbf{Text content} \\
\midrule
Original text &
\begin{CJK}{UTF8}{gbsn}
故事的开头是这样的。一个小男孩，在森林里发现了一个神秘的宝箱，里面装满了金子。
\end{CJK} \\
\midrule
Marked text &
\begin{CJK}{UTF8}{gbsn}
比较慢地说 \texttt{<|endofprompt|>} 故事的开头是这样${\scriptstyle\leftarrow}$的${\scriptstyle\cdot\cdot}$一个小${\scriptstyle\nearrow}$男孩，在${\scriptstyle\uparrow}$森林里发${\scriptstyle\rightarrow}$现了一个${\scriptstyle\nearrow}$神秘${\scriptstyle\rightarrow}$的宝箱…里面装满了金${\scriptstyle\rightarrow}$子。
\end{CJK} \\
\bottomrule
\end{tabular}
\end{table}

\textbf{Pitch-extraction reliability.} Because alignment quality directly
controls marker correctness, we ensemble two aligners and filter by
agreement (above). For pitch we compared PYIN~\cite{mauch2014},
CREPE~\cite{kim2018}, and RMVPE~\cite{wei23rmvpe} and adopted RMVPE for its
superior accuracy--cost trade-off; we acknowledge that ensembling multiple
pitch trackers, analogously to the alignment stage, is a worthwhile
extension.

\subsection{Acoustic Interpreter: Teaching a Frozen-Architecture TTS-LLM to Follow Markers}
\label{subsec:interpreter}

The core of the framework is the observation that the LLM inside an existing
TTS system can be fine-tuned to interpret the injected markers as explicit,
time-aligned prosodic targets, mapping the discrete symbolic sequence to
corresponding acoustic token sequences. Training pairs are produced entirely
by the automatic pipeline of Section~\ref{subsec:modelling}: the input is
marked text and the target is the acoustic token sequence of the
corresponding utterance. The base architecture---tokenizer, LLM, and
decoder---is kept intact; only the input distribution changes. Prosody
conditioning therefore enters through the text stream, which makes the
approach applicable to deployed LLM-based TTS systems in principle, although
our empirical validation is limited to a single backbone
(Section~\ref{sec:experiments}).

\subsection{Optional Natural-Language Front-End}
\label{subsec:frontend}

For usability, markers can also be generated from free-form instructions
rather than extracted from reference audio. We construct triplets
$\{\text{text}, \text{prompt}, \text{marked text}\}$ by \emph{inverse prosody
captioning} with a multimodal LLM (Qwen3-Omni~\cite{xu2025qwen3omni}) and
fine-tune a text LLM (Qwen3-8B~\cite{yang2025qwen3}, LoRA, one epoch,
learning rate $10^{-4}$) to map descriptions to marked text. The training
set comprises approximately 210{,}000 instances curated from three scenarios
with distinct prosodic profiles---TV dramas (60.75\%), lectures (34.97\%),
and live streaming (4.29\%)---spanning seven emotion categories, and is
split 7:2:1 into training, validation, and test partitions. On the held-out
test set, fine-tuning reduces the normalized edit distance (NED;
character-level Levenshtein distance~\cite{levenshtein1966binary} normalized
by reference length and averaged per sequence) between generated and
reference marked text from 7.91 (zero-shot) to 4.61, where reference
annotations are produced by the automatic pipeline of
Section~\ref{subsec:modelling}. We stress that this front-end is an optional
usability layer: the contributions of this paper do not depend on it, and a
full end-to-end instruction-to-speech evaluation is left to future work.

\section{Experiments}
\label{sec:experiments}

We instantiate the framework on CosyVoice2~\cite{zhang2025cosyvoice2} as a
representative LLM-based TTS, chosen for its fully open-source training and
inference stack. The experiments validate (1) whether automatically extracted
markers are learned by an existing TTS-LLM, and (2) whether the resulting
conditioning enables character-level manipulation while perceptual quality is
preserved. All baselines labeled \emph{Base} refer to the unmodified,
publicly released CosyVoice2 model.

\subsection{Dataset and Setup}
\label{subsec:setup}

\textbf{Dataset.} The Mandarin subset of VoxBox~\cite{wang25sparktts} is
processed with the pipeline of Section~\ref{subsec:modelling}. Utterances of
3--25\,s and at most 60 characters are retained, and samples without any
prosodic marker are filtered out, yielding 2{,}150{,}447 utterances (5{,}858
hours). We re-split the retained corpus 9:1 into training and validation
sets, stratified by speaker so that no speaker appears in both splits; the
split seeds and file lists will be released upon publication.

\textbf{Configuration.} Long/short characters are defined as $t>430$\,ms and
$30<t<50$\,ms. Speaking-rate levels (mean character duration) are:
\emph{very slow} ($>275.22$\,ms), \emph{slow} ($223.75$--$275.22$\,ms),
\emph{normal} ($167.82$--$223.75$\,ms), \emph{fast}
($147.4$--$167.82$\,ms), and \emph{very fast} ($<147.4$\,ms). Pitch labeling
follows Section~\ref{subsec:modelling} ($\pm100$\,Hz level shift, $<35$\,Hz
internal range, duration $>150$\,ms).

\textbf{Training.} The TTS-LLM is fine-tuned on 8$\times$B200 GPUs for 20
epochs (Adam, $\beta=(0.9,0.999)$, lr $=10^{-5}$, linear warmup with cosine
annealing). The three checkpoints with the lowest validation loss are fused
by uniform weight averaging before evaluation.

\subsection{Character-Level Controllability}
\label{subsec:controllability}

We synthesize 11 manually composed test sentences designed to cover all
marker types, with target characters manually annotated, across 12 speaker
identities, using both the base model and the marker-conditioned model. F0
and duration of the target characters are measured with the same extraction
stack used at training time. We acknowledge that reusing the training-time
extractor for evaluation could mask systematic extraction errors; however,
since base and conditioned models are scored identically, any such bias
affects both conditions equally, and the \emph{relative} changes in
Table~\ref{tab:control} remain meaningful. Absolute calibration against
human annotation is left to future work.

\textbf{Duration and pitch.} Table~\ref{tab:control} summarizes the results
(mean per condition). The marker-conditioned model lengthens target
characters by 62.9\% (240\,$\rightarrow$\,391\,ms) and shortens them by
17.4\% (144\,$\rightarrow$\,119\,ms); rising markers increase the F0 shift
from $+6$ to $+11$\,Hz and falling markers increase the downward shift in
magnitude from $8$ to $12$\,Hz ($+50\%$). We report means over all
speaker--sentence pairs; distributional analysis is identified as a
limitation in Section~\ref{sec:discussion}. Note that, by design, this
evaluation applies lengthen/shorten markers to characters that are already
relatively long or short at baseline, measuring the additional shift induced
by the marker rather than an absolute target duration.

\begin{table}[t]
\centering
\caption{Character-level prosody controllability (means over 12 speakers
$\times$ 11 sentences).}
\label{tab:control}
\begin{tabular}{@{}lccc@{}}
\toprule
\textbf{Control} & \textbf{Base} & \textbf{Ours} & \textbf{Change} \\
\midrule
Lengthen (ms) & 240 & 391 & +62.9\% \\
Shorten (ms) & 144 & 119 & --17.4\% \\
Rising F0 shift (Hz) & +6 & +11 & +83.3\% \\
Falling F0 shift (Hz) & --8 & --12 & +50.0\% \\
\bottomrule
\end{tabular}
\end{table}

\textbf{Pause duration.} We synthesize 2{,}000 samples per pause-marker
type, inserting each marker at inter-phrase positions (sentence-medial
punctuation boundaries) in held-out text. This evaluates the realized pause
duration conditioned on the marker, independent of placement prediction.
Measured pauses (Table~\ref{tab:pause_duration}) fall within the target
ranges used during training.

\begin{table}[t]
\centering
\caption{Pause controllability: target ranges vs.\ measured means
(2{,}000 samples per marker).}
\label{tab:pause_duration}
\begin{tabular}{ccc}
\toprule
Symbol & Target range (ms) & Measured mean (ms) \\
\midrule
,              & [300, 500)   & 354  \\
$\cdot\cdot$   & [500, 800)   & 525  \\
…              & [800, 1000)  & 824  \\
$\circ$        & [1000, 1500) & 1109 \\
\bottomrule
\end{tabular}
\end{table}

\begin{figure}[t]
\centering
\includegraphics[width=\linewidth]{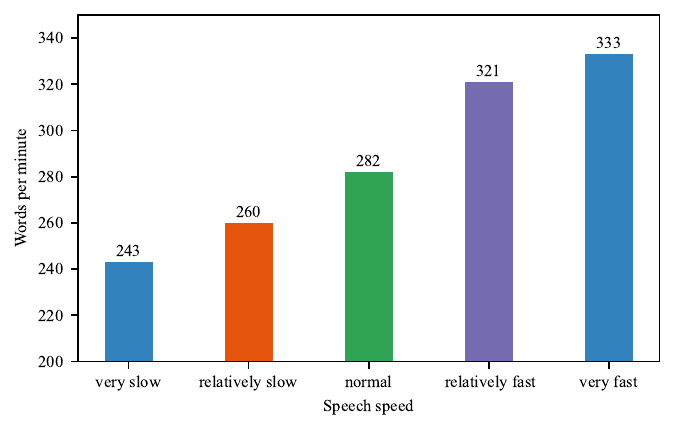}
\caption{Mean per-utterance character duration conditioned on the five
global rate prompts (group means across speakers).}
\Description{Grouped bar or line chart of mean character duration for the
five speaking-rate prompts, showing monotone separation of group means.}
\label{fig:speed}
\end{figure}

\textbf{Global speaking rate.} The five rate levels are defined a priori
from corpus percentiles (Section~\ref{subsec:setup}); the evaluation
question is therefore \emph{whether the model reproduces distinct rate
behavior when conditioned on each prompt}, not whether the levels themselves
exist. Figure~\ref{fig:speed} reports the mean per-utterance character
duration grouped by rate prompt across speakers: group means are well
separated and ordered consistently with the prompt, with no systematic
interaction with speaker identity. We note that the figure summarizes group
means; we therefore refrain from claiming non-overlapping distributions on
the basis of means alone, and leave variance-level analysis to future work.

\textbf{Energy.} We do not report loudness-control results: the tokenizer
analysis below (Section~\ref{subsec:tokenizer}) shows that the backbone's
ASR-optimized tokens are largely invariant to energy perturbations, and the
measured effects were marginal.

\subsection{Perceptual Evaluation}
\label{subsec:perceptual}

We conducted a listening study with 24 listeners (volunteers, all
self-reported native Mandarin speakers with normal hearing). Each listener
rated 24 stimuli (12 per system, randomly interleaved and balanced across
listeners) on 5-point scales along three dimensions: \emph{naturalness},
\emph{realism} (prosodic plausibility and similarity to the original
utterance), and \emph{speaker similarity}. Reference recordings of the
original utterances were provided during the realism and speaker-similarity
blocks; no MOS anchor stimuli were used. The study took approximately 20
minutes per participant.

Table~\ref{tab:mos} reports mean scores on emotionally charged content. The
marker-conditioned model tracks the base model closely on naturalness (4.44
vs.\ 4.46) and scores marginally higher on realism (4.15 vs.\ 4.10) and
speaker similarity (4.00 vs.\ 3.95). All differences are within fractions of
a scale point; we report them \emph{descriptively} and do not claim
statistical significance, as the study was not powered or analyzed for
inferential testing. Recent methodological analyses of MOS-based
evaluation~\cite{kirkland2023stuck,cooper2023investigating,lemaguer2024limits,camp2023mos}
recommend comparative protocols (AB/CMOS, BWS) over absolute MOS for small
system differences~\cite{wells2024experimental}; adopting such a protocol,
together with confidence intervals, is planned follow-up work.

\begin{table}[t]
\centering
\caption{Perceptual evaluation (mean opinion scores, 24 listeners;
descriptive comparison on emotionally charged content; Base = unmodified
CosyVoice2).}
\label{tab:mos}
\begin{tabular}{@{}lcc@{}}
\toprule
\textbf{Attribute} & \textbf{Base} & \textbf{Ours} \\
\midrule
Naturalness        & 4.46 & 4.44 \\
Realism            & 4.10 & 4.15 \\
Speaker similarity & 3.95 & 4.00 \\
\bottomrule
\end{tabular}
\end{table}

\subsection{Cross-Tokenizer Sensitivity Analysis}
\label{subsec:tokenizer}

\begin{figure}[t]
\centering
\includegraphics[width=\linewidth]{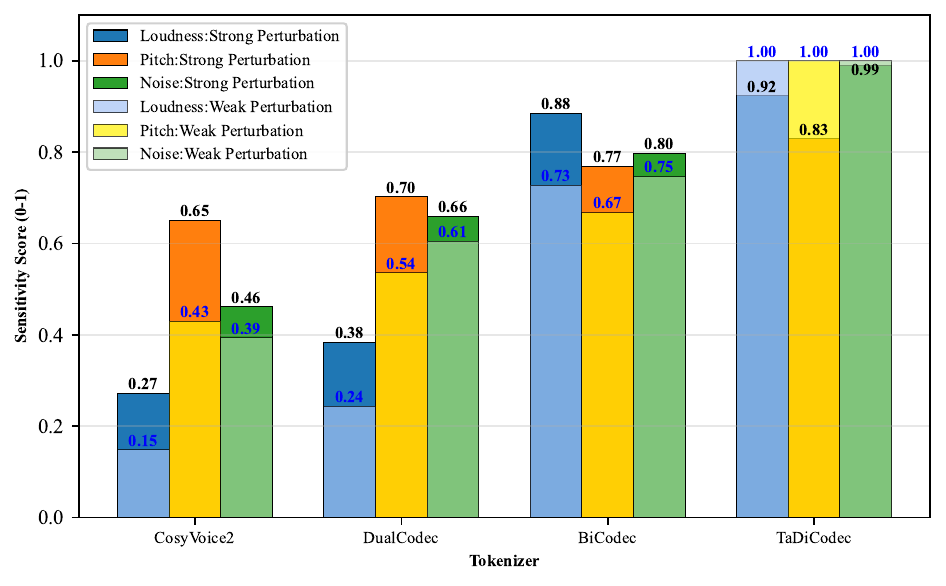}
\caption{Tokenizer sensitivity to acoustic perturbations (NED between
original and perturbed token sequences; higher = more sensitive). CosyVoice2
(ASR-optimized) is least sensitive overall; reconstruction-oriented
tokenizers are more sensitive, with DualCodec lying in between, suggesting a
continuum rather than a strict dichotomy.}
\Description{Bar chart of normalized edit distance between original and
perturbed token sequences for four tokenizers across volume, F0, and noise
perturbations.}
\label{fig:tokenizer}
\end{figure}

\textbf{Setup.} We perturb 1{,}000 audio samples (volume $\pm5/10$\,dB, F0
$\pm1/3$ semitones, additive noise at $-45/-40$\,dB relative level, i.e.,
SNR $\approx+45/+40$\,dB) and measure the NED (normalized edit
distance~\cite{levenshtein1966binary}) between the original and perturbed
token sequences for four tokenizers: CosyVoice2 (ASR-optimized) and the
reconstruction-oriented BiCodec~\cite{wang25sparktts},
DualCodec~\cite{du2025dualcodec}, and TaDiCodec~\cite{wang2025tadicodec}.

\textbf{Results.} CosyVoice2 exhibits the lowest NED across all
perturbations, and is particularly insensitive to energy
variation---consistent with our decision not to report loudness control.
TaDiCodec nearly rewrites the token sequence under strong perturbations
(NED$\approx$1.0), while DualCodec lies between the extremes. Because only
one ASR-optimized tokenizer is included and the pattern is graded rather
than categorical, we present these results as \emph{hypothesis-generating}:
they are consistent with a trade-off in which ASR objectives favor
linguistic content at the expense of prosodic detail, but they do not by
themselves demonstrate such a mechanism.

\textbf{Relation to the backbone choice.} This observation does not
contradict the controllability results above: our markers inject prosodic
targets through the \emph{text} stream, and the fine-tuned LLM learns to
realize them despite tokenizer-level attenuation. Rather, the analysis
suggests that reconstruction-oriented tokenizers may offer larger headroom
for marker-based control---a direction we leave to future work.

\section{Limitations and Future Work}
\label{sec:discussion}

\textbf{Language and backbone scope.} Validation is limited to Mandarin
Chinese and a single backbone (CosyVoice2). Character-level alignment does
not transfer directly to non-tonal or agglutinative languages, and extension
to other tokenizers and architectures---reconstruction-oriented codecs such
as EnCodec~\cite{defossez2023high} or SoundStream~\cite{zeghidour2022soundstream},
and non-LLM architectures---remains future work.

\textbf{Marker design.} Thresholds are fixed constants; relative or
log-scale/semitone F0 thresholds, and ensembling of pitch trackers, may
improve robustness. Extraction quality depends on forced-alignment accuracy
and may degrade for spontaneous or atypical speech.

\textbf{Evaluation scope.} The perceptual study is descriptive (24
listeners, no inferential testing) and controllability results are reported
as condition means; comparative protocols with confidence intervals and
distributional analysis are needed. Marker quality was not compared against human
annotation, and the optional instruction front-end was evaluated only at the
text level, not end-to-end. Evaluation of the TTS-LLM uses a held-out
validation partition; a fully independent test set with disjoint recording
conditions is left to future work. Strong marker combinations can trade small amounts of neutral naturalness for realism in expressive contexts; quantifying this trade-off with comparative listening tests is planned.

\section{Conclusion}
\label{sec:conclusion}

We presented a data-driven framework for character-level prosody control of
Mandarin LLM-based TTS. Prosodic markers are extracted automatically from
raw audio and embedded as Unicode symbols, and an existing TTS-LLM learns to
interpret them through fine-tuning alone, providing fine-grained temporal
control complementary to utterance-level methods without architectural
changes. Experiments at the scale of 2.15M utterances demonstrate reliable
control of duration, pitch, pauses, and speaking rate, and a cross-tokenizer
analysis offers hypothesis-generating evidence that ASR-optimized tokenizers
attenuate prosodic information. Future work will extend validation across
languages and backbones, adopt comparative perceptual protocols, and
evaluate the full instruction-to-markers-to-speech pathway end to end.

\section*{Ethics and Privacy Statement}
Fine-grained prosody control lowers the effort needed to produce expressive
synthetic speech, which could be misused for deceptive or impersonating
audio content. Our experiments use a publicly released corpus and
open-source models, and the listening study involved volunteer raters
evaluating synthetic stimuli. We advocate pairing such control techniques
with synthetic-speech disclosure and detection mechanisms in deployed
multimedia applications.

\bibliographystyle{ACM-Reference-Format}
\bibliography{references}

@article{du24cosyvoice,
  author = {Du, Zhihao and Chen, Qian and Zhang, Shiliang and Hu, Kai and Lu, Heng and Yang, Yexin and others},
  title = {CosyVoice: A Scalable Multilingual Zero-Shot Text-to-Speech Synthesizer Based on Supervised Semantic Tokens},
  journal = {arXiv preprint arXiv:2407.05407},
  year = {2024}
}

@article{wang25sparktts,
  author = {Wang, Xinsheng and Jiang, Mingqi and Ma, Ziyang and Zhang, Ziyu and Liu, Songxiang and Li, Linqin and others},
  title = {Spark-TTS: An Efficient LLM-Based Text-to-Speech Model with Single-Stream Decoupled Speech Tokens},
  journal = {arXiv preprint arXiv:2503.01710},
  year = {2025}
}

@article{zeng2024seedtts,
  author = {Anastassiou, Philippos and Chen, Jiawei and Chen, Jitong and Chen, Yuanzhe and Chen, Zhuo and Chen, Ziyi and others},
  title = {Seed-TTS: A Family of High-Quality Versatile Speech Generation Models},
  journal = {arXiv preprint arXiv:2406.02430},
  year = {2024}
}

@article{wang2023neural,
  author = {Chen, Sanyuan and Wang, Chengyi and Wu, Yu and Zhang, Ziqiang and Zhou, Long and Liu, Shujie and others},
  title = {Neural Codec Language Models Are Zero-Shot Text to Speech Synthesizers},
  journal = {IEEE/ACM Transactions on Audio, Speech, and Language Processing},
  volume = {33},
  pages = {705--718},
  year = {2025}
}

@article{deng25indextts,
  author = {Deng, Wei and Zhou, Siyi and Shu, Jingchen and Wang, Jinchao and Wang, Lu},
  title = {IndexTTS: An Industrial-Level Controllable and Efficient Zero-Shot Text-to-Speech System},
  journal = {arXiv preprint arXiv:2502.05512},
  year = {2025}
}

@article{zhang2025cosyvoice2,
  author = {Du, Zhihao and Wang, Yuxuan and Chen, Qian and Shi, Xian and Lv, Xiang and Zhao, Tianyu and others},
  title = {CosyVoice 2: Scalable Streaming Speech Synthesis with Large Language Models},
  journal = {arXiv preprint arXiv:2412.10117},
  year = {2024}
}

@inproceedings{xie2025towards,
  author = {Xie, Tianxin and Rong, Yan and Zhang, Pengfei and Wang, Wenwu and Liu, Li},
  title = {Towards Controllable Speech Synthesis in the Era of Large Language Models: A Systematic Survey},
  booktitle = {Proceedings of the 2025 Conference on Empirical Methods in Natural Language Processing (EMNLP)},
  pages = {764--791},
  year = {2025}
}

@article{yang2024instructtts,
  author = {Yang, Dongchao and Liu, Songxiang and Huang, Rongjie and Weng, Chao and Meng, Helen},
  title = {InstructTTS: Modelling Expressive TTS in Discrete Latent Space with Natural Language Style Prompt},
  journal = {IEEE/ACM Transactions on Audio, Speech, and Language Processing},
  volume = {32},
  pages = {2913--2925},
  year = {2024}
}

@inproceedings{leng24prompttts,
  author = {Leng, Yichong and Guo, Zhifang and Shen, Kai and Tan, Xu and Ju, Zeqian and Liu, Yanqing and others},
  title = {PromptTTS 2: Describing and Generating Voices with Text Prompt},
  booktitle = {Proceedings of the Twelfth International Conference on Learning Representations (ICLR)},
  year = {2024}
}

@inproceedings{shimizu24prompttts,
  author = {Shimizu, Reo and Yamamoto, Ryuichi and Kawamura, Masaya and Shirahata, Yuma and Doi, Hironori and Komatsu, Tatsuya and others},
  title = {PromptTTS++: Controlling Speaker Identity in Prompt-Based Text-to-Speech Using Natural Language Descriptions},
  booktitle = {Proc. ICASSP},
  pages = {12672--12676},
  year = {2024}
}

@inproceedings{liu23promptstyle,
  author = {Liu, Guanghou and Zhang, Yongmao and Lei, Yi and Chen, Yunlin and Wang, Rui and Li, Zhifei and others},
  title = {PromptStyle: Controllable Style Transfer for Text-to-Speech with Natural Language Descriptions},
  booktitle = {Proc. Interspeech},
  pages = {4888--4892},
  year = {2023}
}

@inproceedings{jiang24megatts2,
  author = {Jiang, Ziyue and Liu, Jinglin and Ren, Yi and He, Jinzheng and Ye, Zhenhui and Ji, Shengpeng and others},
  title = {Mega-TTS 2: Boosting Prompting Mechanisms for Zero-Shot Speech Synthesis},
  booktitle = {Proceedings of the Twelfth International Conference on Learning Representations (ICLR)},
  year = {2024}
}

@inproceedings{shen24naturalspeech2,
  author = {Shen, Kai and Ju, Zeqian and Tan, Xu and Liu, Yanqing and Leng, Yichong and He, Lei and others},
  title = {NaturalSpeech 2: Latent Diffusion Models Are Natural and Zero-Shot Speech and Singing Synthesizers},
  booktitle = {Proceedings of the Twelfth International Conference on Learning Representations (ICLR)},
  year = {2024}
}

@inproceedings{ju24naturalspeech3,
  author = {Ju, Zeqian and Wang, Yuancheng and Shen, Kai and Tan, Xu and Xin, Detai and Yang, Dongchao and others},
  title = {NaturalSpeech 3: Zero-Shot Speech Synthesis with Factorized Codec and Diffusion Models},
  booktitle = {Proceedings of the 41st International Conference on Machine Learning (ICML)},
  pages = {22605--22623},
  year = {2024}
}

@inproceedings{wang2018style,
  author = {Wang, Yuxuan and Stanton, Daisy and Zhang, Yu and Skerry-Ryan, R. J. and Battenberg, Eric and Shor, Joel and others},
  title = {Style Tokens: Unsupervised Style Modeling, Control and Transfer in End-to-End Speech Synthesis},
  booktitle = {Proceedings of the 35th International Conference on Machine Learning (ICML)},
  pages = {5180--5189},
  year = {2018}
}

@inproceedings{wu2024laugh,
  author = {Wu, Haibin and Wang, Xiaofei and Eskimez, Sefik Emre and Thakker, Manthan and Tompkins, Daniel and Tsai, Chung-Hsien and others},
  title = {Laugh Now Cry Later: Controlling Time-Varying Emotional States of Flow-Matching-Based Zero-Shot Text-to-Speech},
  booktitle = {Proceedings of the IEEE Spoken Language Technology Workshop (SLT)},
  pages = {690--697},
  year = {2024}
}

@article{lei22msemotts,
  author = {Lei, Yi and Yang, Shan and Wang, Xinsheng and Xie, Lei},
  title = {MsEmoTTS: Multi-Scale Emotion Transfer, Prediction, and Control for Emotional Speech Synthesis},
  journal = {IEEE/ACM Transactions on Audio, Speech, and Language Processing},
  volume = {30},
  pages = {853--864},
  year = {2022}
}

@inproceedings{silverman1992tobi,
  author = {Silverman, Kim and Beckman, Mary and Pitrelli, John and Ostendorf, Mari and Wightman, Colin and Price, Patti and others},
  title = {TOBI: A Standard for Labeling English Prosody},
  booktitle = {Proceedings of the 2nd International Conference on Spoken Language Processing (ICSLP)},
  pages = {867--870},
  year = {1992}
}

@inproceedings{ren2020fastspeech,
  author = {Ren, Yi and Hu, Chenxu and Tan, Xu and Qin, Tao and Zhao, Sheng and Zhao, Zhou and others},
  title = {FastSpeech 2: Fast and High-Quality End-to-End Text to Speech},
  booktitle = {Proceedings of the Ninth International Conference on Learning Representations (ICLR)},
  year = {2021}
}

@inproceedings{korotkova2024wordlevel,
  author = {Korotkova, Yuliya and Kalinovskiy, Ilya and Vakhrusheva, Tatiana},
  title = {Word-Level Text Markup for Prosody Control in Speech Synthesis},
  booktitle = {Proc. Interspeech},
  pages = {2280--2284},
  year = {2024}
}

@inproceedings{guo2022unsupervised,
  author = {Guo, Yiwei and Du, Chenpeng and Yu, Kai},
  title = {Unsupervised Word-Level Prosody Tagging for Controllable Speech Synthesis},
  booktitle = {Proc. ICASSP},
  pages = {7597--7601},
  year = {2022}
}

@inproceedings{mohan2021ctrlp,
  author = {Mohan, Devang S. and others},
  title = {Ctrl-P: Temporal Control of Prosodic Variation for Speech Synthesis},
  booktitle = {Proc. Interspeech},
  pages = {3875--3879},
  year = {2021}
}

@inproceedings{vioni2021prosodic,
  author = {Vioni, Alexandra and others},
  title = {Prosodic Clustering for Phoneme-Level Prosody Control in End-to-End Speech Synthesis},
  booktitle = {Proc. ICASSP},
  pages = {5719--5723},
  year = {2021}
}

@inproceedings{yoshimura1999simultaneous,
  author = {Yoshimura, Takayoshi and Tokuda, Keiichi and Masuko, Takashi and Kobayashi, Takao and Kitamura, Tadashi},
  title = {Simultaneous Modeling of Spectrum, Pitch and Duration in HMM-Based Speech Synthesis},
  booktitle = {Proc. Eurospeech},
  pages = {2347--2350},
  year = {1999}
}

@inproceedings{dai2022automatic,
  author = {Dai, Ziqian and Yu, Jianwei and Wang, Yan and Chen, Nuo and Bian, Yanyao and Li, Guangzhi and Cai, Deng and Yu, Dong},
  title = {Automatic Prosody Annotation with Pre-Trained Text-Speech Model},
  booktitle = {Proc. Interspeech},
  pages = {5513--5517},
  year = {2022}
}

@inproceedings{tilson2024toward,
  author = {Tilson, Alex and F{\"o}rster, Frank},
  title = {Toward Real Time Word Based Prosody Recognition},
  booktitle = {Proceedings of the CLASP Conference on Multimodality and Interaction in Language Learning},
  year = {2024}
}

@inproceedings{wang25batonvoice,
  author = {Wang, Yue and Ma, Ruotian and Chen, Xingyu and Shi, Zhengliang and Yang, Morunliu and Chen, Wanshun and Liu, Huang and Yao, Jiadi and He, Xin and Yang, Qu and Jiang, Qingxuan and Ye, Fanghua and Li, Juntao and Tu, Zhaopeng and Li, Xiaolong and Bo, Liefeng and Zhang, Min},
  title = {BatonVoice: An Operationalist Framework for Enhancing Controllable Speech Synthesis with Linguistic Intelligence from LLMs},
  booktitle = {Proceedings of the 64th Annual Meeting of the Association for Computational Linguistics (ACL)},
  pages = {46683--46697},
  year = {2026},
  note = {Published version; see \url{https://aclanthology.org/2026.acl-long.2165/}}
}

@article{liao2025nvspeech,
  author = {Liao, Huan and Ni, Qinke and Wang, Yuancheng and Lu, Yiheng and Zhan, Haoyue and Xie, Pengyuan and Zhang, Qiang and Wu, Zhizheng},
  title = {NVSpeech: An Integrated and Scalable Pipeline for Human-Like Speech Modeling with Paralinguistic Vocalizations},
  journal = {arXiv preprint arXiv:2508.04195},
  year = {2025}
}

@inproceedings{mcauliffe17_interspeech,
  author = {McAuliffe, Michael and Socolof, Michaela and Mihuc, Sarah and Wagner, Michael and Sonderegger, Morgan},
  title = {Montreal Forced Aligner: Trainable Text-Speech Alignment Using Kaldi},
  booktitle = {Proc. Interspeech},
  pages = {498--502},
  year = {2017}
}

@inproceedings{bain23_interspeech,
  author = {Bain, Max and Huh, Jaesung and Han, Tengda and Zisserman, Andrew},
  title = {WhisperX: Time-Accurate Speech Transcription of Long-Form Audio},
  booktitle = {Proc. Interspeech},
  pages = {4489--4493},
  year = {2023}
}

@inproceedings{graves2006ctc,
  author = {Graves, Alex and Fern{\'a}ndez, Santiago and Gomez, Faustino and Schmidhuber, J{\"u}rgen},
  title = {Connectionist Temporal Classification: Labelling Unsegmented Sequence Data with Recurrent Neural Networks},
  booktitle = {Proceedings of the 23rd International Conference on Machine Learning (ICML)},
  pages = {369--376},
  year = {2006}
}

@inproceedings{wei23rmvpe,
  author = {Wei, Haojie and Cao, Xueke and Dan, Tangpeng and Chen, Yueguo},
  title = {RMVPE: A Robust Model for Vocal Pitch Estimation in Polyphonic Music},
  booktitle = {Proc. Interspeech},
  pages = {5421--5425},
  year = {2023}
}

@inproceedings{mauch2014,
  author = {Mauch, Matthias and Dixon, Simon},
  title = {PYIN: A Fundamental Frequency Estimator Using Probabilistic Threshold Distributions},
  booktitle = {Proc. ICASSP},
  pages = {659--663},
  year = {2014}
}

@inproceedings{kim2018,
  author = {Kim, Jong Wook and Salamon, Justin and Li, Peter and Bello, Juan Pablo},
  title = {CREPE: A Convolutional Representation for Pitch Estimation},
  booktitle = {Proc. ICASSP},
  pages = {161--165},
  year = {2018}
}

@inproceedings{du2025dualcodec,
  author = {Li, Jiaqi and Lin, Xiaolong and Li, Zhekai and Huang, Shixi and Wang, Yuancheng and Wang, Chaoren and Zhan, Zhenpeng and Wu, Zhizheng},
  title = {DualCodec: A Low-Frame-Rate, Semantically-Enhanced Neural Audio Codec for Speech Generation},
  booktitle = {Proc. Interspeech},
  pages = {4883--4887},
  year = {2025}
}

@inproceedings{wang2025tadicodec,
  author = {Wang, Yuancheng and Chen, Dekun and Zhang, Xueyao and Zhang, Junan and Li, Jiaqi and Wu, Zhizheng},
  title = {TaDiCodec: Text-Aware Diffusion Speech Tokenizer for Speech Language Modeling},
  booktitle = {Advances in Neural Information Processing Systems (NeurIPS)},
  year = {2025}
}

@article{defossez2023high,
  author = {D{\'e}fossez, Alexandre and Copet, Jade and Synnaeve, Gabriel and Adi, Yossi},
  title = {High Fidelity Neural Audio Compression},
  journal = {Transactions on Machine Learning Research},
  year = {2023}
}

@article{zeghidour2022soundstream,
  author = {Zeghidour, Neil and Luebs, Alejandro and Omran, Ahmed and Skoglund, Jan and Tagliasacchi, Marco},
  title = {SoundStream: An End-to-End Neural Audio Codec},
  journal = {IEEE/ACM Transactions on Audio, Speech, and Language Processing},
  volume = {30},
  pages = {495--507},
  year = {2022}
}

@article{xu2025qwen3omni,
  author = {Xu, Jin and Guo, Zhifang and Hu, Hangrui and Chu, Yunfei and Wang, Xiong and He, Jinzheng and others},
  title = {Qwen3-Omni Technical Report},
  journal = {arXiv preprint arXiv:2509.17765},
  year = {2025}
}

@article{yang2025qwen3,
  author = {Yang, An and Li, Anfeng and Yang, Baosong and Zhang, Beichen and Hui, Binyuan and Zheng, Bo and others},
  title = {Qwen3 Technical Report},
  journal = {arXiv preprint arXiv:2505.09388},
  year = {2025}
}

@article{levenshtein1966binary,
  author = {Levenshtein, Vladimir I.},
  title = {Binary Codes Capable of Correcting Deletions, Insertions, and Reversals},
  journal = {Soviet Physics Doklady},
  volume = {10},
  number = {8},
  pages = {707--710},
  year = {1966}
}

@inproceedings{kirkland2023stuck,
  author = {Kirkland, Ambika and Mehta, Shivam and Lameris, Harm and Henter, Gustav Eje and Sz{\'e}kely, {\'E}va and Gustafson, Joakim},
  title = {Stuck in the MOS Pit: A Critical Analysis of MOS Test Methodology in TTS Evaluation},
  booktitle = {Proceedings of the 12th ISCA Speech Synthesis Workshop (SSW)},
  pages = {41--47},
  year = {2023}
}

@inproceedings{cooper2023investigating,
  author = {Cooper, Erica and Yamagishi, Junichi},
  title = {Investigating Range-Equalizing Bias in Mean Opinion Score Ratings of Synthesized Speech},
  booktitle = {Proc. Interspeech},
  pages = {1104--1108},
  year = {2023}
}

@article{lemaguer2024limits,
  author = {Le Maguer, S{\'e}bastien and King, Simon and Harte, Naomi},
  title = {The Limits of the Mean Opinion Score for Speech Synthesis Evaluation},
  journal = {Computer Speech \& Language},
  volume = {84},
  pages = {101577},
  year = {2024}
}

@inproceedings{camp2023mos,
  author = {Camp, Joshua and Kenter, Tom and Finkelstein, Lev and Clark, Rob},
  title = {MOS vs. AB: Evaluating Text-to-Speech Systems Reliably Using Clustered Standard Errors},
  booktitle = {Proc. Interspeech},
  pages = {1090--1094},
  year = {2023}
}

@inproceedings{wells2024experimental,
  author = {Wells, Dan and Aldana Blanco, Andrea Lorena and Valentini, Cassia and Cooper, Erica and Pine, Aidan and Yamagishi, Junichi and Richmond, Korin},
  title = {Experimental Evaluation of MOS, AB and BWS Listening Test Designs},
  booktitle = {Proc. Interspeech},
  pages = {2695--2699},
  year = {2024}
}

\end{document}